\documentclass{nolta}

\usepackage[dvipdfmx]{graphicx}
\usepackage{dcolumn}
\usepackage{bm}
\usepackage{amsmath}
\usepackage{amssymb}
\usepackage{url}
\Vol{X}%
\No{0}%
\Year{2023}%
\Month{99}%

\received{12}{06}{2022}

\title{Time series analysis using persistent homology of distance matrix}

\AUTHOR{
\author[a]{Takashi Ichinomiya}{1,2}\orcid{0000-0002-9173-4514}
}

\AFFILIATE{
    \affiliate{%
     Gifu University School of Medicine,\\
    Yanagido 1-1, Gifu, Gifu 501-1194, Japan
    }{1}%
    \affiliate{
    The United Graduate School of Drug Discovery and Medical Informatic Science, Gifu University,\\ 
    Yanagido 1-1, Gifu 501-1194, Japan
    }{2}
}
\EMAIL{
\email{tk1miya@gifu-u.ac.jp}{a}
}

\begin{document}

\begin{abstract}
 The analysis of nonlinear dynamics is an important issue in numerous fields of science.  In this study, we propose a new method to analyze the time series data using persistent homology (PH). The key idea is the application of PH to the distance matrix. Using this method, we can obtain the topological features embedded in the trajectories. We apply this method to the logistic map, R\"ossler system, and electrocardiogram data. The results reveal that our method can effectively identify nonlocal characteristics of the attractor and can classify data based on the amount of noise.
\end{abstract}

\begin{keywords}
 time-series analysis, persistent homology, dynamical system
\end{keywords}

\maketitle


\section{Introduction}
 The analysis of the nonlinear dynamics is a challenging problem in physics, engineering, and data science. Numerous methods for time series analysis,  such as Fourier transformation or Kalman filtering, are based on the theory of linear dynamics, and the performance of these methods is limited. To overcome their limitations, several methods, such as Koopman's mode decomposition\cite{Budisic2012AppliedKoopmanism}, phase reduction\cite{Nakao2016PhaseOscillators}, deep learning\cite{IsmailFawaz2019DeepReview}, and reservoir computing\cite{Jaeger2004HarnessingCommunication,Maass2002Real-TimePerturbations}, have been proposed. However, we often meet the situations where these methods are not available. In Koopman's mode decomposition we map the finite-dimensional nonlinear dynamical system into an infinite-dimensional linear dynamical system, and investigate the eigenvalues and eigenfunctions in this space. However, the determination of these eigenfunctions is often difficult. 
 Phase reduction is a powerful method to investigate the oscillatory dynamics, but the application to non-oscillatory systems is limited. Deep learning and reservoir computing help us to predict the state in the future, but they do not provide a rationale for their prediction.
 
 In this study, we propose a new method to analyze the dynamics using distance matrix. The distance matrix $D(t, s)$, defined as the distance between states at time $t$ and $s$, reveals considerable information regarding the dynamics of the system. For example, if $D(t,t+T)=0$ for all $t$, the system exhibits a periodic motion with period $T$. Recurrence plot (RP) is the most useful method for the time-series analysis using a distance matrix \cite{Eckmann1987}. In this method, the distance matrix is visualized using $R(t, s) = \Theta (\epsilon-D(t,s))$, where $\Theta(x)$ is the Heaviside step function and $\epsilon$ is a parameter. Using this method, periodic motion can be distinguished from chaotic trajectories. Based on an RP, recurrence quantification analysis (RQA), wherein the dynamics are characterized by several quantities, such as recurrence rate and determinism, was proposed.
 
 However, an RP uses only the limited information embedded in the distance matrix. First, an RP does not provide information on the nonlocal properties of the trajectory. An RP lists the points that are close to each other in phase space and is useful to investigate local properties such as Lyapunov exponents. In contrast, it is unsuitable for investigating the global structure of the trajectory. For example, determining whether the attractor has a double scroll structure like the Lorenz system or an oscillation-like structure similar to the R\"ossler system by RP is difficult. Another problem with the use of an RP is that there is no clear rule to select the value of $\epsilon$, and studies have reported that the result of the RP is often sensitive to this choice \cite{Marwan2007}.
 
 In this study, we propose the application of persistent homology (PH)\cite{Edelsbrunner2002}, an emerging technique of data analysis, to the analysis of the distance matrix. PH is the one of the most popular techniques in topological data analysis (TDA). In TDA, we investigate the topological characteristics such as number of connected components or holes embedded in the dataset. The theory of PH is still being developed and has been successfully applied in various fields, including biophysics\cite{Xia2014,Xia2015,Ichinomiya2020}, material science\cite{Kimura2018,Ichinomiya2017,Hiraoka2016}, and image processing\cite{Letscher2007ImagePersistence,Qaiser2019FastFeatures}. 

There have been several proposals to apply PH to time series datasets. The favored approach for the time-series analysis using PH is based on delay embedding \cite{Perea2015,Maletic2016}. In this approach, we first create the point clouds in $n$-dimensional space using Takens' delay embedding \cite{Takens1981} and subsequently characterize the state using PH. However, this approach has several difficulties. First, we must determine how to embed the dataset. There is no general rule to determine the way of embedding, though several ideas have been proposed \cite{Cellucci2003ComparativeMethods,Deyle2011GeneralizedReconstruction,Jia2020RefinedSeries}.  Second, the computational cost increases rapidly as the embedding dimension and the size of point cloud increases. It is known that the computation time for PH is $O(N^{\lceil D/2 \rceil})$, where $N$ and $D$ represent the size of the point cloud and dimension of the space, respectively \cite{Otter2017AHomology}. Therefore, the computational cost increases exponentially as $D$ increases.
In our approach, we can avoid the latter difficulty. Because the distance matrix is represented in two-dimensional space, the computation cost of PH is considerably reduced. The results of this study reveal that using PH, the essential information of a dataset, such as the non-local structure of attractors and the amount of noise, can be easily determined.

The remainder of this study is structured as follows. In Section ~\ref{Sec:method}, we explain our method to investigate the distance matrix. In Section ~\ref{Sec:result}, we present the results of application of our method to three different datasets. First, we present the results of the analysis of the logistic map as a typical example of discrete time dynamics. Second, we present the results on the R\"ossler system as an example of continuous time dynamics. Third,  we discuss the results on the analysis of electrocardiogram (ECG) data, as an example of real-world time series data. Finally, we discuss on the possible improvements to our method in future and conclude this study.

\section{Method}\label{Sec:method}

 The general definition of PH requires further background knowledge of algebraic topology. In this section, we explain the PH of filtered cubical complexes of degree 0, which is used in the latter part of this study. The readers who want to know the general definition of PH can consult textbooks on PH and TDA\cite{Edelsbrunner2010,Carlsson2021TopologicalApplications}.

\begin{figure}
    \centering
    \includegraphics[width=.7\textwidth]{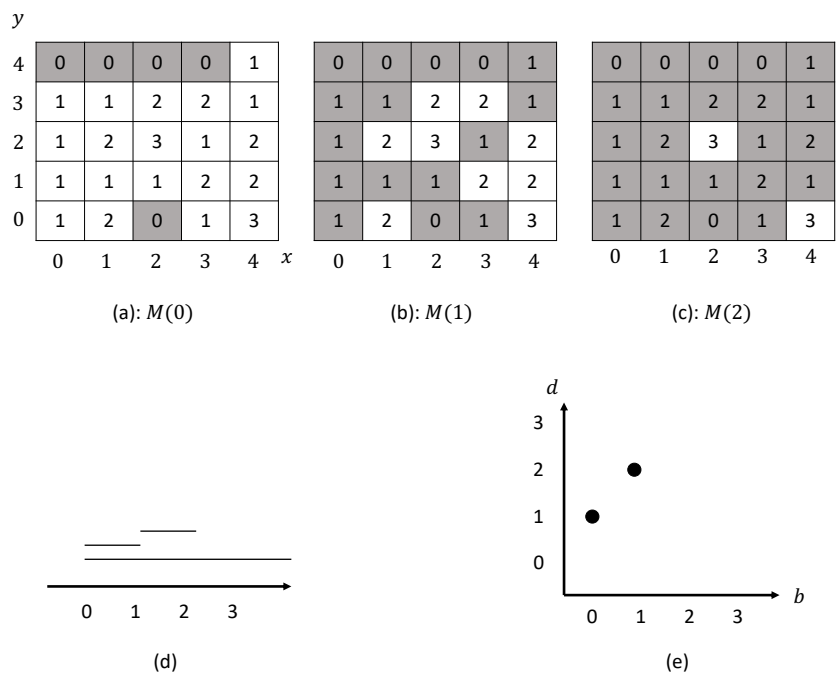}
    \caption{Example of persistent homology using sublevel sets, (a)--(c) represent the $M(\theta)$ for $\theta=0, 1$, and 2, respectively. The corresponding persistence barcode and persistence diagram are shown in (d) and (e).}
    \label{fig:sublevel}
\end{figure}

 We consider a real-valued filtration function $f:\mathbb{Z}^2\rightarrow \mathbb{R}$. The sublevel set $M(\theta)$ is defined by
 \begin{equation}
     M(\theta) =\{(x,y)\in \mathbb{Z}^2 | f(x,y) \le \theta \}.
 \end{equation}
For example, assume that $f(x,y)$ is given by the table shown in Fig.~\ref{fig:sublevel}(a). In this case, $M(0)$ is given by the gray blocks. When we increase $\theta$, $M(\theta)$ also grows, as shown in Fig.~\ref{fig:sublevel}(b) and (c). 

PH with degree 0 using a sublevel set of $f$ represents the change in connected components in $M(\theta)$ when $\theta$ is varied from $-\infty$ to $\infty$. Here we say two blocks are ``connected'' if they share an edge. For example, we have two components in Fig.~\ref{fig:sublevel}(a): there is one isolated component at $(x,y) = (2,0)$ and one connected component at $y=4, 0\le x \le 3$.  By increasing $\theta$ to 1, these two components are merged into one large component, and another component appears at $(x,y)=(3,2)$, as shown in Fig.~\ref{fig:sublevel}(b). Here, we do not say these two components are connected, because although they share two corners, they do not share an edge. In this case, we say these two components are ``disconnected.''
When we increase $\theta$ to 2, all three components are merged into one large component, as shown in Fig.~\ref{fig:sublevel}(c). The theory of PH guarantees that we can define the ``birth'' and ``death'' of each component. In this example, at $\theta=0$, two disconnected components are ``born.'' At $\theta=1$, these components merged into one large component. Hence, we say that one of the components ``dies'' and the other disconnected component at $(x,y) = (3,2)$ is born. Finally, at $\theta=2$, these components are connected. Further increase in $\theta$ does not change the number of disconnected components. Therefore, we have three ``connected components,'' often called ``generators,'' whose births $b$ and deaths $d$ are $(b,d)= (0,\infty), (0,1), (1,2)$, respectively.

 There are two major visualization techniques to represent the distribution of generators. One is ``persistence barcode,'' wherein we represent each generator as a ``bar'' from birth to death, as shown in Fig.~\ref{fig:sublevel}(d). This representation is intuitive when the number of generators is small. For example, when the dataset has a periodic structure, we will obtain numerous generators that have the same birth and death, and we can easily identify them by persistence barcode. However, when we have more than a hundred of generators, the number of bars becomes too large, and gaining any insight from the barcode becomes difficult. In this case, a persistence diagram (PD) is better visualization method; herein, we make a scatter plot of birth and death, as shown in Fig.~\ref{fig:sublevel}(e). In a PD, generators with infinite death are generally omitted. When the number of generators becomes large, we also use a density heatmap of generators, which is also called a PD. In the rest of this study, we use PDs to represent the results of our PH analysis. 
 
\begin{figure}
    \centering
    \includegraphics[width=.6\textwidth]{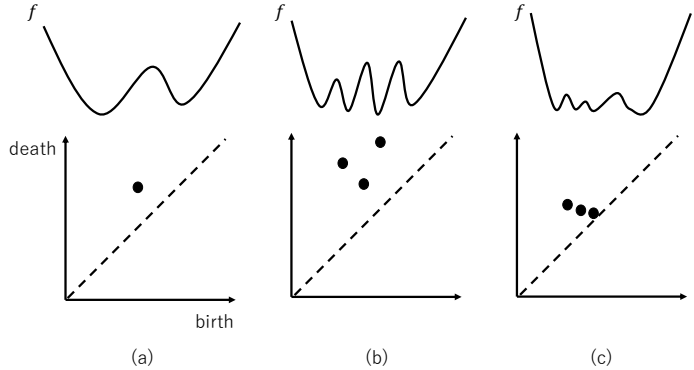}
    \caption{Relation between the form of the filtering function $f$ (upper) and corresponding persistence diagram(lower). (a) When $f$ has two local minima, we obtain two generators, and only one has finite death. (b) When $f$ is more complex and has more local minima, the number of generators increases. (c) If the ``saddles'' between local minima are low, the lifetime of generators decreases. The dashed line in the persistence diagrams indicates the line birth = death. }
    \label{fig:strucuture_and_pd}
\end{figure}

The PH allows us to study the local minimum and saddle points of $f$. For example, we consider the case of Fig.~\ref{fig:strucuture_and_pd}. In the case of Fig.~\ref{fig:strucuture_and_pd}(a), $f$ is a smooth function with two minima. In this case, we can define the sublevel set $M(\theta)$ as $M(\theta) = \{x \in \mathbb{R} | f(x) \le \theta\}$. If $\theta$ is smaller than the saddle value of $f$, $M(\theta)$ has two connected components, and for larger $\theta$, $M(\theta)$ has only one connected component. Therefore, in this case, there are two generators, and one of these generators has finite death. In contrast, if $D$ has numerous local minima as described in Fig.~\ref{fig:strucuture_and_pd}(b), we obtain numerous generators with finite deaths. Therefore, the number of generators indicates the number of local minima of $f$. Moreover, PH provides information about the height of the saddles. For example, we consider the case in Fig.~\ref{fig:strucuture_and_pd}(c). Here, $f$ has numerous local minima, but the height of the saddle is lower than in Fig.~\ref{fig:strucuture_and_pd}(b). In this case, the connected components of $M(\theta)$ merge after only a slight increase of $\theta$. Therefore, the lifetime, defined as the difference between death and birth, decreases as the height of the corresponding saddle decreases. 

In this study, we used a distance matrix $D$ for filtration. We suppose that we have time series data $\boldsymbol{x}_i$, $i=1,2,\ldots, K$, where $i$ represents the discretized time. From this dataset, we define the distance matrix $D(i,j)$ as 
\begin{equation}
    D(i,j) = \lvert\lvert \boldsymbol{x}_i-\boldsymbol{x}_j \lvert \lvert,
\end{equation}
where $\lvert \lvert \cdots \lvert \lvert$ represents $L^2$-norm.
$D(i,j)$ is a real-valued function from $(i,j)\in \mathbb{Z}^2$, and we can apply PH using the distance matrix. 

There are several software for PH analysis, which include Gudhi \cite{Maria2014TheHomology}, Phat \cite{Bauer2017PhatToolbox}, and Javaplex \cite{Javaplex}. In this study, we used Homcloud developed by Obayashi {\it et al} \cite{Obayashi2022PersistentHomCloud}. One of the advantages of Homcloud compared with other software is that it can provide the ``birth position'' and ``death position'' of each generator. Using this function, we can obtain the positions of the local minima and the saddles of $D(i,j)$. These values are useful for interpreting the result obtained by PH.

\section{Results}\label{Sec:result}

We applied our method to three different datasets. The first example is time-series data obtained from the logistic map, and the second one is that obtained from the R\"ossler equations. Finally, we analyzed the dataset ECG200, as an example of a real-world dataset.

\subsection{Analysis of the logistic map}

In this subsection, we investigated the distance matrix of the logistic map defined by
\begin{equation}
    x_{i+1} =a x_i (1-x_i),
\end{equation}
where $0 < a < 4$ is a parameter. We calculated $x_i$ for $i=1,2, \ldots, 1000$ with initial condition $x_0=0.832$ and performed a PH analysis using data at $i=801,802, \ldots, 1000$. 



\begin{figure}
    \centering
    \includegraphics[width=.8\textwidth]{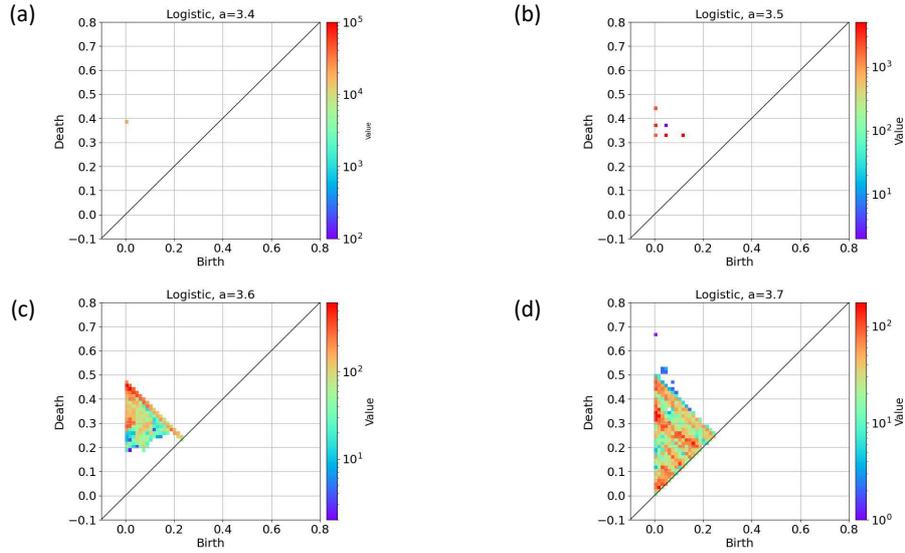}
    \caption{Persistence diagram for logistic maps: (a) $a=3.4$, (b) $a=3.5$, (c) $a=3.6$, and (d) $a=3.7$. }
    \label{fig:Logistic_PD}
\end{figure}

\begin{figure}
 \centering
 \includegraphics[width=0.8\textwidth]{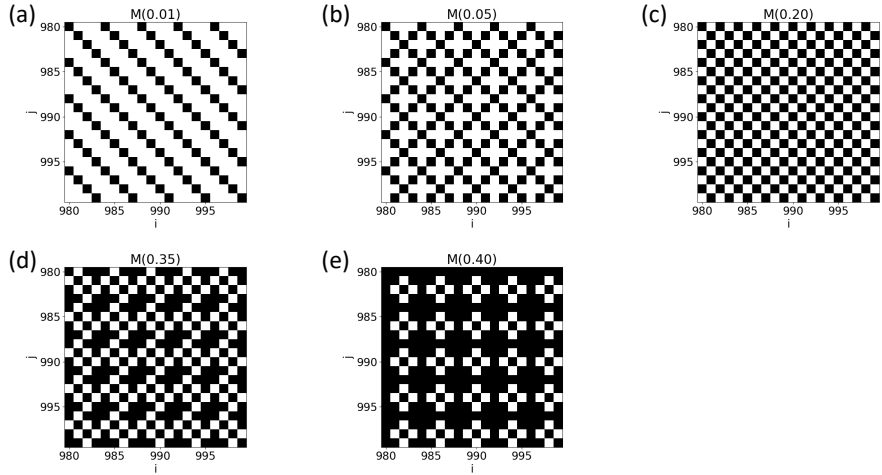}
\caption{Sublevel set $M(\theta)$ represented as black areas in the case of $a=3.5$. (a) $\theta=0.01$, (b) $\theta=0.05$, (c) $\theta=0.20$, (d) $\theta=0.35$, and (e) $\theta=0.40$, respectively.}
\label{fig:Logistic_sublevel}
\end{figure}

First, we investigated the case $a=3.4$, wherein $x_i$ is periodic, shown in Fig.~\ref{fig:Logistic_PD}(a). In this case, all generators had birth 0 and death 0.3901. This result indicated that $x_i$ takes only two values, and the difference between these two is 0.3901. This is consistent with the fact that the attractor of this system is a cycle with period 2: $x_{2i+1}= 0.4520$ and $x_{2i} = 0.8421$. The death time is given by the difference between these two values. As the period increases, the number of values that generators can take also increases. For example, at $a=3.5$, the distribution had 6 peaks, corresponding to the fact that the period of the logistic map was 4. We show $M(\theta)$ for several $\theta$ in Fig.\ref{fig:Logistic_sublevel} in the case of $a=3.5$. The number of peaks represents the topological information of the attractors.

\begin{figure}
    \centering
    \includegraphics[width=.45\textwidth]{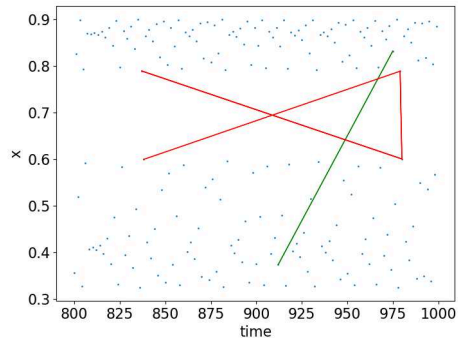}
    \caption{Time series $x_i$ for the case of $a=3.6$. Red lines indicate the death position of generators whose lifetime is smaller than 0.19. The green line indicates the death position of the generators with the largest birth+death, birth = 0.0104 and death = 0.4590.}
    \label{fig:birth_position}
\end{figure}

Furthermore, the PD in a chaotic region provides topological information of the attractor. In the case of $a=3.6$, where the logistic map becomes chaotic, the births and deaths of generators are widely distributed, as shown in Fig.~\ref{fig:Logistic_PD}(c). Herein, we found that the distribution had several characteristic properties. First, all deaths were larger than 0.188. This property could be explained by the existence of a gap in the attractor. At $a=3.6$, $x$ takes values from 0.3 to 0.6 and from 0.788 to 0.900, but does not take values between 0.601 to 0.787. This property generates a the gap in the distribution of deaths. To confirm this suggestion, we show the ``death point'' of generators whose death is smaller than 0.19 as a red line in Fig.~\ref{fig:birth_position}. In this figure, the points connected by lines give the saddles of $D(i,j)$. Evidently, the red lines connect the top point of the lower band and the bottom points of the upper band, which supports our claim.

Additionally, Fig.~\ref{fig:Logistic_PD}(c) reveals that birth + death $< 0.4695$. Unfortunately, we have no theory to explain this property. Notably, we have an upper limit on birth + death when $x_{min} \le x_i \le x_{max}$ for all $i$, because birth $\ge 0$ and death $\le x_{max}-x_{min}$. However, the fact that the death corresponds to the saddle makes the problem difficult. For example, the green line in Fig.~\ref{fig:Logistic_PD}(e) shows the death point pairs with the largest birth + death, 0.4694.  This figure shows that one terminal of this line is at the upper limit of the upper band, whereas the other terminal stays in the middle of the lower band. We have no theoretical explanation why there is no saddle pair of states that provides larger birth + death. This is a problem that should be solved in the future.

In the case of $a=3.7$ shown in Fig.~\ref{fig:Logistic_PD}(d), the upper limit of death+birth appears to remain with several exceptional generators, whereas the lower limit disappears. The absence of a lower limit indicates the elimination of the gap in $x$ found in the case $a=3.6$. Summarizing these results, the PD gives the topological structure of the attractor, in the cases of both the periodic trajectory and chaotic attractor.

\subsection{Analysis of the R\"ossler system}

Next, we investigated the dynamics of the R\"ossler system described by the equations
\begin{align}
    \frac{dx}{dt} & = -y -z \\
    \frac{dy}{dt} & = x + a y \\
    \frac{dz}{dt} & = b + xz -cz .
\end{align}
In this study, we set $a=0.2, c=5.7$, and investigated the change in the PD by varying $b$ between 0.1 and 1.7. We calculated $(x(t), y(t), z(t))$ for $0 \le t \le 500$, and used $(x(t), y(t), z(t))$ for $t=400, 400.1,\ldots, 499.9$ for the calculation of the distance matrix. 

\begin{figure}
    \centering
    \includegraphics[width=.7\textwidth]{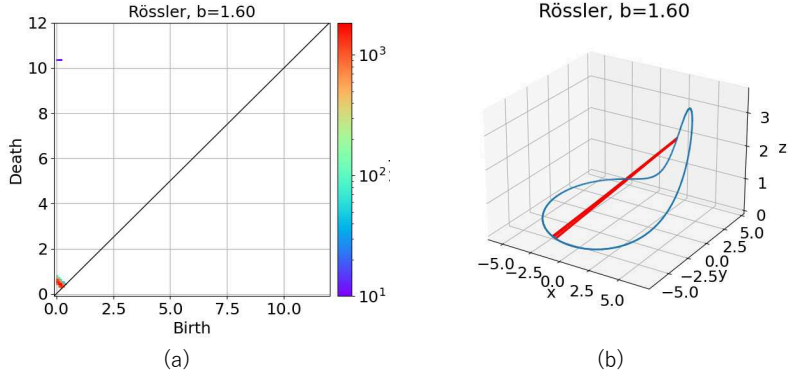}
    \caption{(a) Persistence diagram, and (b) trajectory of R\"ossler system, with $a=0.2, b=1.6, c=5.7$. Red lines in (b) represent the death positions of generators whose deaths are larger than 10.0.}
    \label{fig:rossler1.6}
\end{figure}

First, we began from the case with $b=1.6$. In this case, the PD shown in Fig.~\ref{fig:rossler1.6}(a) shows that there are two classes of generators: generators in the first group have deaths smaller than 1.0 and those in the second group have deaths larger than 10.0. To study the origin of the generators with large death, we investigated the ``death position'' of these generators, which are indicated by the red lines in Fig.~\ref{fig:rossler1.6}(b). At this parameter, the attractor was the orbit with period 1, and the large death value indicated the ``diameter'' of this orbit.

\begin{figure}
    \centering
    \includegraphics[width=.7\textwidth]{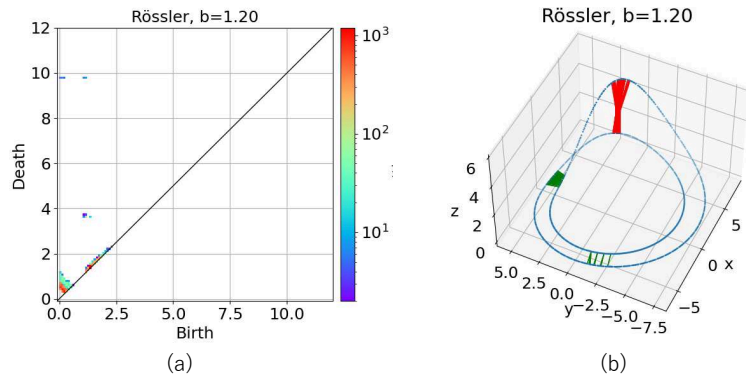}
    \caption{(a): Persistence diagram, and (b) trajectory of R\"ossler system, $a=0.2, b=1.2, c=5.7$. Green and red lines in Fig. (b) represent the birth and death positions of generators whose deaths are between 3.0 and 4.0, respectively.}
    \label{fig:rossler1.2}
\end{figure}

\begin{figure}
    \centering
    \includegraphics[width=.7\textwidth]{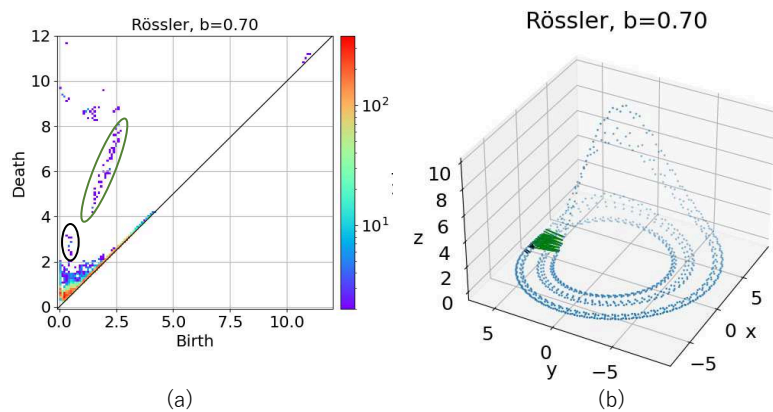}
    \caption{(a) Persistence diagram, and (b) trajectory of R\"ossler system, $a=0.2, b=0.7, c=5.7$. The birth positions of generators surrounded by black and green ellipses in (a) correspond to the black and green lines in (b), respectively.}
    \label{fig:rossler0.7}
\end{figure}

Next, we demonstrated the PD for $b=1.2$ in Fig.~\ref{fig:rossler1.2}(a). The PD was similar to the case of $b=1.6$, but new generators whose deaths are approximately 4.0 appeared. These generators represent the period doubling of the attractor. To reveal the relation between period doubling and the generators, we show the trajectories of the system with the birth and death positions of these generators in Fig.~\ref{fig:rossler1.2}. Evidently, the birth positions of these generators, which are represented by green lines, connect the parallel part of the trajectory. These generators dies at the red lines, where these two lines diverge along the $z$ axis. This example suggests that the generators with large births and deaths imply the existence of a parallel separated trajectory in the attractor. This claim holds true in the chaotic region. For example, we show the PD at $b=0.7$ in Fig.~\ref{fig:rossler0.7}(a); herein, the system had a chaotic attractor. The PD shows several clusters of generators with large lifetimes, and these generators represent the parallel trajectories. For example, the birth of the generators surrounded by black and green ellipses in Fig.~\ref{fig:rossler0.7}(a) correspond with the black and green lines in Fig.~\ref{fig:rossler0.7}(b), respectively. These birth points are the pairs between parallel trajectories in the phase space.
These examples suggest that we can identify the nonlocal structure of the attractors using PH.

\begin{figure}
 \centering
 \includegraphics[width=.8\textwidth]{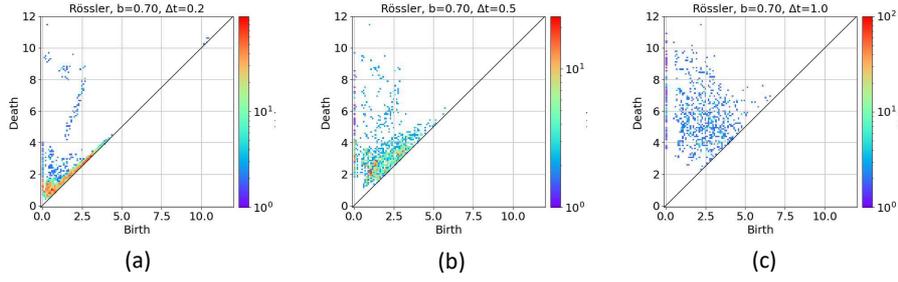}
 \caption{Persistence diagrams for several interval of snapshots $\Delta t$. (a) $\Delta t= 0.2$, (b) $\Delta t=0.5$, and (c)$\Delta t = 1.0$, respectively. The pararmeters of the R\"ossler system are $a=0.2, b=0.7, c=5.7$}
 \label{fig:rossler_dt}
\end{figure}

\begin{figure}
 \centering
 \includegraphics[width=.7\textwidth]{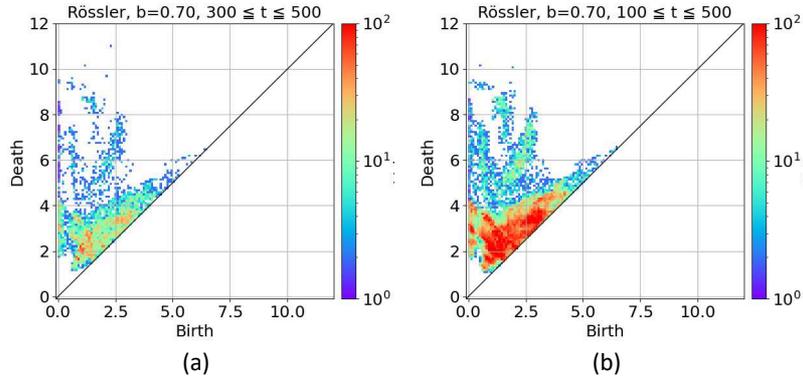}
 \caption{Persistence diagrams obtained using the  trajectory (a) $300\le t \le 500$, and  (b) $100\le t \le 500$. The parameters of the R\"ossler system are $a=0.2, =0.7, c=5.7$, and the interval of snapshots $\Delta t = 0.5$.}
 \label{fig:rossler_length}
\end{figure}

In the analysis above, we analyzed the dynamics using the snapshots of the system with an interval $\Delta t= 0.1$. It is natural to ask whether our results are robust against the change of the interval. We calculated the PDs for several $\Delta t$, using the trajectory with $b=0.7$, $400\le t \le 500$. Here, we notice that the number of snapshots decreases as $\Delta t$ increases. The result is shown in Fig.~\ref{fig:rossler_dt}. When $\Delta t=0.2$, we obtained the PD shown in Fig.~\ref{fig:rossler_dt}(a), which is qualitatively consistent with the case of $\Delta t= 0.1$ in Fig.~\ref{fig:rossler0.7}(a). Further increase of  $\Delta t $ produced the many ``noisy'' generators, which made the structure of generators unclear. Fig.~\ref{fig:rossler_dt} (b) and (c) represent the PDs where $\Delta t= 0.5$ and 1.0, respectively. In the case of $\Delta t = 0.5$, the PD has many noisy generators, but it seems similar to the PD with $\Delta t = 0.1$ and 0.2.  In the case of $\Delta t = 1.0$, we find no clear cluster of generators.
These ``noisy'' generators cannot be removed even if we use a longer time series. Fig.~\ref{fig:rossler_length} shows the PDs when we take longer sequences for $\Delta t = 0.5$. Fig.~\ref{fig:rossler_length}(a) and (b) use two times and four times longer sequences than the case of Fig.~\ref{fig:rossler_dt}(b), but the ``noisy'' generators remain. These results suggests that a small $\Delta t$ is required to apply our method.
 

\subsection{Analysis of the ECG200 dataset}

In this subsection, we present the analysis of the real time-series dataset ECG200, provided by Olszewski {\it et al}\cite{Olszewski2001GeneralizedData}. This dataset includes the 200 ECGs of a heartbeat, which are classified into two classes: healthy patients and patients with a myocardial infarction (MI). The dataset is divided into 100 training samples and 100 test samples. In this study, we only used the dataset for training. The dataset was downloaded from the Time Series Classification Repository\cite{Bagnall2017TheAdvances}. 

\begin{figure}
    \centering
    \includegraphics[width=.8\textwidth]{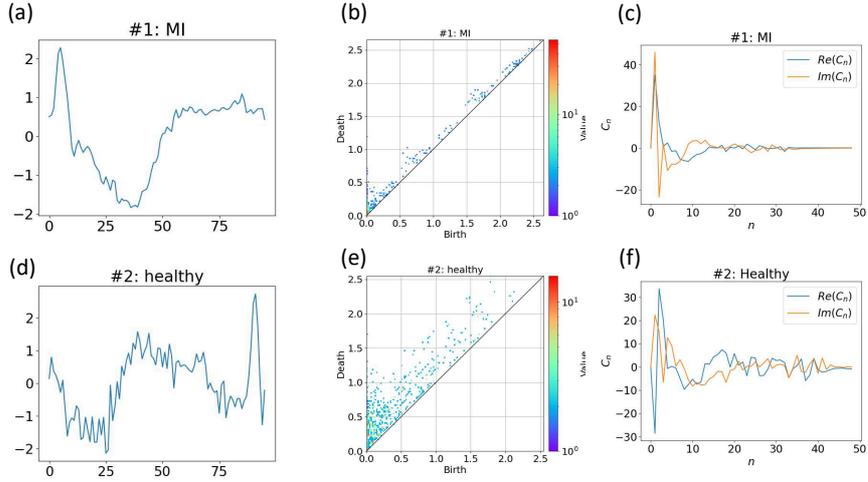}
    \caption{Examples of electrocardiogram data and corresponding persistence diagrams and Fourier components. Upper: (a) electrocardiogram data, (b) persistence diagram, and (c) Fourier components obtained from a healthy patients.  Lower: (d) electrocardiogram data, (e) persisntence diagram, and (f) Fourier components obtained from a patient with myocardial infarction. }
    \label{fig:ecgsample}
\end{figure}

First, we present the examples of ECG data of a healthy patient and a patient with an MI in Figs.~\ref{fig:ecgsample} (a) and (d). In this figure, the ECG signal of an MI patient appears flat, whereas that of the healthy patient contains considerable noise. The corresponding PDs are shown in Figs.~\ref{fig:ecgsample}(b) and (e). In the case of an MI patient shown in Fig.~\ref{fig:ecgsample} (e), the lifetime of most generators was smaller than 0.5. In contrast, the lifetimes in Fig.~\ref{fig:ecgsample}(e) had a large variation, which suggested the existence of noise in the data from a healthy patient. In this case, the PH gives more intuition to us than the standard technique such as Fourier transformation. For example, we present the results of Fourier transformation $c_n = \sum_m x(m) \exp\left(\frac{2\pi i m n}{N}\right)$ in Fig.~\ref{fig:ecgsample}(c) and (f), where $N$ represents the length of sequence. The difference of the coefficients between the healthy patient and the patient with MI seems clear, but it is difficult to define a single variable that classify these two classes. Fourier transformation is a powerful tool when the time series has some characteristic frequencies, but in this case, there is no typical frequency that distinguishes patients with MI from healthy patients. To apply Fourier transformation in this problem, we require more complicated methods such as the analysis using Bag-of-SFA-Symbols\cite{Schafer2015}. 

\begin{figure}
    \centering
    \includegraphics[width=.65\textwidth]{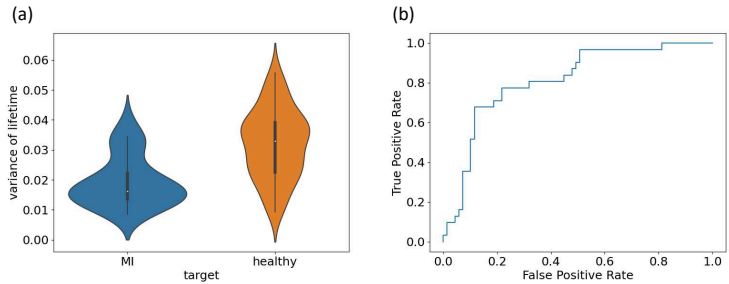}
    \caption{(a) Distribution of variance of lifetimes for MI patients and healthy patients. (b) Receiver operating curve when we use variance of lifetime as the indicator of MI. Area under curve = 0.811. }
    \label{fig:ecgstat}
\end{figure}

Based on this observation, we investigated whether we can use the variance in the lifetimes of the generators as an indicator for an MI. First, we studied the distribution of the variance of lifetimes for patients with MI and healthy patients. The result is shown in Fig.~\ref{fig:ecgstat}(a). In the case of an MI, the distribution peaked around 0.01, whereas in the case of the healthy patients, it peaked around 0.04 for normal persons. This figure suggests that low variance in lifetimes is the signal of an MI. To estimate the performance of the variance of lifetimes as the marker of an MI, we calculated the receiver operating characteristic (ROC) curve and the area under the curve (AUC). The ROC curve represents the relation between false positive rate (FPR) and true positive rate (TPR). Suppose that we judge the ECG whose variance of lifetimes is smaller than $p$ is positive. Then, number of true positive (TP), false negative (FN), true negative (TN), and false positive (FP) are defined as the number of correctly identified MI patients, misidentified MI patients, correctly identified healthy patients, and misidentified healthy patients, respectively. TPR and FPR are defined by
\begin{equation}
    TPR =\frac{TP}{TP+FN},
\end{equation}
and
\begin{equation}
    FPR=\frac{FP}{FP+TN}.
\end{equation}

The ROC curve is the plot of (FPR, TPR). AUC, defined as the area under the ROC curve, is the standard characteristic of the performance of a quantitative diagnostic test. If AUC = 1.0, we have no incorrect identification, whereas if AUC=0.5, the identification is equivalent to a random identification.
In our case, the AUC was 0.811, which implies that the variance has a moderate accuracy as an indicator of MI \cite{Bowers2019ReceiverOutcomes}. 
Compared with this result, Kirchenko {\it et al.} classified the same dataset using the deep learning of RQA and the RP image, with AUC=0.76 and 0.92, respectively \cite{Kirichenko2021ApplyingSeries}.  Therefore, our method is better than analysis using RQA, but it does not improve upon the analysis of the RP using deep learning. However, we note that the interpretation of our result is much easier than that obtained by deep learning.
In summary, our example shows that the variance of lifetimes is a promising signal to diagnose MI.

\section{Discussion and Conclusion}\label{Sec:conclusion}

In this study, we proposed a new method for time-series analysis, using PH analysis of the distance matrix. We demonstrated the efficacy of our method to understand the structure of attractors in the logistic map and the R\"ossler systems; additionally, we demonstrated that this method is applicable to real-world datasets such as ECG200. Our method uses the distance matrix, which is represented in two-dimensional space, thereby saving computational cost compared with other methods. Thus, our method is useful for analyzing dynamics in high-dimensional systems.

However, PH is not a developed method for data analysis, and there remains room for improvement in our method. To conclude this study, we discuss several directions for future studies.

First, combining this technique with machine learning is a promising approach. To analyze a large real-world dataset, machine learning techniques, such as deep learning, must be used. In machine learning, vectorizing the characteristic features is essential. However, the number of generators produced by PH is not constant, thereby rendering difficulty in the application of standard machine learning techniques such as principal component analysis. Moreover, the generators with small lifetime are often disregarded as meaningless because they are produced by small noise, and the ``significance'' of each generator must be estimated. To overcome these difficulties, numerous researchers have proposed a variety of techniques such as persistence landscape \cite{Bubenik2015}, persistence images \cite{Adams2017PersistenceHomology}, and persistence weighted Gaussian kernel \cite{Kusano2016}. These techniques combined with our proposed method will enable the application of our method to numerous problems.

Second, investigating the information embedded in different PDs is interesting. In this study, we did not use the PDs of degree 1, which provide characteristics of ``holes'' surrounded by $M(\theta)$. The generators with degree 1 provide insights on the peaks of the distance matrix, which may give us essential insights on the dynamics.  Additionally, we can also calculate PDs using dilation-erosion filtration \cite{Garin2019ATDA}. In this approach, first, we make a recurrence plot, and subsequently calculate the distance to the ``boundary,'' the places where black and white cells contact, for each cell. PH analysis using this ``distance to the boundary'' as a filtration function provides a new quantification of the recurrence plot. However, to generate a PD with dilation-erosion, we must determine the threshold for the recurrence plot. The multi-parameter PH, which is currently studied intensively \cite{Carlsson2009ThePersistence,Vipond2020MultiparameterLandscapes.}, is the PH method with several filtration functions. The application of this method will provide a way to combine our method and dilation-erosion based PH analysis. 

Finally, another future task is to mathematically investigate the relation between PD and dynamics. In the case of RQA, it is known that the determinism has relation to the positive Lyapunov exponents. It would be an interesting to seek the mathematical relation between the PDs and the characteristics of the dynamical systems.


\section{Acknowledgemens}

This work is financially supported by 	JSPS KAKENHI Grant Number JP22K19816.

\bibliography{references}

\end{document}